# Small-signal model for 2D-material based field-effect transistors targeting radio-frequency applications: the importance of considering non-reciprocal capacitances

Francisco Pasadas, Wei Wei, Emiliano Pallecchi, Henri Happy and David Jiménez

*Abstract*— A small-signal equivalent circuit of 2D-material based field-effect transistors is presented. Charge conservation and non-reciprocal capacitances have been assumed so the model can be used to make reliable predictions at both device and circuit levels. In this context, explicit and exact analytical expressions of the main radio-frequency figures of merit of these devices are given. Moreover, a direct parameter extraction methodology is provided based on S-parameter measurements. In addition to the intrinsic capacitances, transconductance and output conductance, our approach allows extracting the series combination of drain/source metal contact and access resistances. Accounting for these extrinsic resistances is of upmost importance when dealing with low dimensional field-effect transistors.

## I. INTRODUCTION

RESEARCH into 2D-material based field-effect transistors (2D-FETs) is propelling the state-of-the-art of digital and high-frequency electronics both on rigid and flexible substrates [1]–[4]. Ongoing efforts are focused on the demonstration of 2D-FETs outperforming the power consumption of Si MOSFETs in digital applications and 2D-FETs working at terahertz frequencies exhibiting power gain. In parallel, there is a great deal of interest in developing digital and radio-frequency (RF) optimized transistors on flexible substrates [5], [6]. A number of advances in those directions have been made in a short time and even a number of simple circuits have been demonstrated [7], [8].

2D-FETs are now operating within the millimeter-wave range showing intrinsic cut-off frequencies ranging from tens to hundreds of gigahertz, and maximum oscillation frequencies up to tens of gigahertz [9]–[11]. Consequently, there is a demand for accurate device models for optimizing the device operation; benchmarking of device performances against other existing technologies; and bridging the gap between device and circuit levels.

In this work, we have developed a small-signal equivalent circuit suited to three-terminal 2D-FETs (see Figs. 1-2). The model formulation is general and applicable to any 2D-material such as graphene and 2D-semiconductors. Different to other previous models that have been applied to 2D-FETs [3], [10]–[14], our model is a charge-based small-signal model, which implies that charge conservation is guaranteed and there is not any unphysical assumption about capacitance reciprocity in the capacitive scheme. Based on such a small-signal model, we have derived explicit expressions for the RF figures of merit (FoMs) with no approximations. We have found discrepancies between the results obtained from our explicit expressions and results obtained from different reported formulas used to evaluate the RF FoMs, especially when a 2D-FET is operated in the negative differential resistance (NDR) region. Finally, a methodology to extract the small-signal parameters from S-parameter measurements is proposed. Importantly, we have included the series combination of the drain/source contact and access resistances to the intrinsic equivalent circuit, which could have a dominant role in the electrical behavior of 2D-FETs. So, our approach allows extracting the source/drain resistance without relying on the use of the transfer length method (TLM) technique, which would imply the fabrication and characterization of devices with different channel lengths [15]. To assess the parameter extraction methodology, we have fed the extracted parameters into the small-signal model and calculated the corresponding S-parameters and RF FoMs. These results have been compared with measurements of an exemplary RF graphene field-effect transistor (GFET).

Manuscript received Month Day, Year; accepted Month Day, Year. Date of publication Month Day, Year; date of current version Month Day, Year. This work was supported in part by the European Union Seventh Framework Programme through the Graphene Flagship Program under Grant 604391, in part by the European Union's through the Horizon 2020 Research and Innovation Programme under Grant 696656, and in part by the Ministerio de Economía y Competitividad under Grant TEC2012-31330 and Grant TEC2015-67462-C2-1-R.

F. Pasadas and D. Jiménez are with the Department of Electronic Engineering, Escola d'Enginyeria, Universitat Autònoma de Barcelona, Bellaterra (Barcelona) 08193, Spain (e-mail: francisco.pasadas@uab.es).

W. Wei, E. Pallecchi and H. Happy are with the Institute of electronics, Microelectronics and Nanotechnology, CNRS UMR8520, Villeneuve d'Ascq 59652, France.

## II. METHODS

### A. Charge-based small-signal equivalent circuit

When considering analog and RF electronic applications, the FET terminals are polarized with a DC bias over which an AC signal is superimposed. The amplitude of the AC signal is usually small enough so the I-V characteristic can be linearized around the DC bias [16]. This way a non-linear device can be treated as a linear circuit with conductance and capacitance elements forming a lumped network.

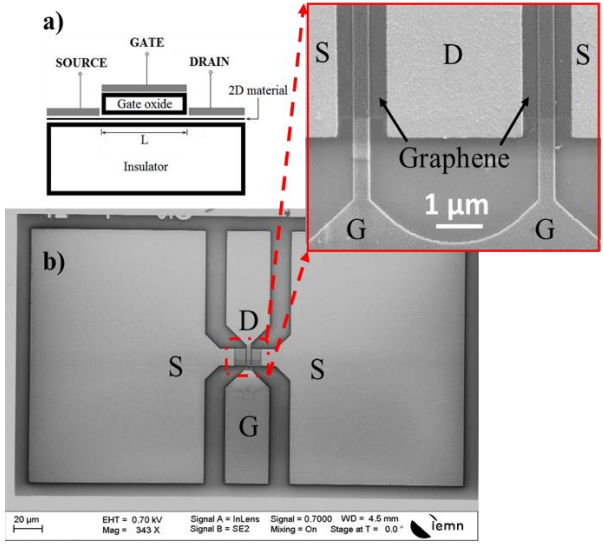

Fig. 1 a) Cross section of a three-terminal 2D-material based field-effect transistor. A 2D-material sheet plays the role of the active channel. The modulation of the carrier population in the channel is achieved via a top-gate stack consisting of a dielectric and corresponding metal gate. b) As an example of a 2D-FET, a scanning electron microscope (SEM) image of the GFET that is considered in section III.B.

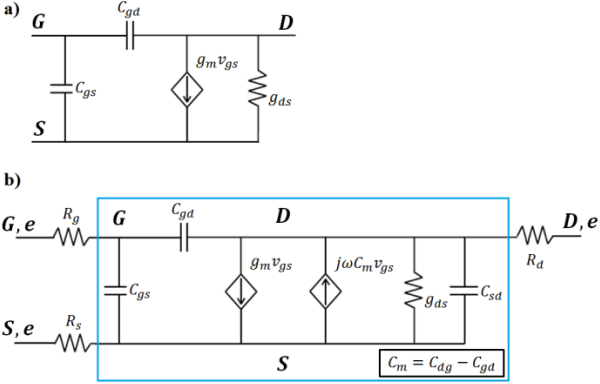

Fig. 2 a) Meyer-like intrinsic small-signal model for a three-terminal FET. b) Charge-based small-signal model suited to 2D-FETs. The equivalent circuit of the intrinsic device is framed in blue. The small-signal elements are: $g_m$ transconductance, $g_{ds}$ output conductance and $C_{gs}$, $C_{gd}$, $C_{sd}$ and $C_{dg}$ intrinsic capacitances. The physical meaning of the elements is explained in [17] for a GFET. $R_g$ is the gate resistance and $R_d$, $R_s$ account for the contact and access resistances of the drain and source respectively.

So far, the small-signal equivalent circuits proposed for 2D-FETs are directly imported from Meyer-like capacitance models [3], [10]–[14]. This kind of models can be represented with the equivalent circuit shown in Fig. 2a. They assume that the intrinsic capacitances of a FET are reciprocal, which is unphysical for a three-terminal device, resulting in important inaccuracies when RF FoMs are evaluated, as we will later show for the case of a GFET. Moreover, these models usually do not ensure charge conservation (although there are exceptions in the literature), which is of upmost importance not only for accurate device modeling and circuit simulation [18]–[22], but even more for proper parameter extraction [23]. In this paper, we propose, instead, the charge-based small-signal model shown in Fig. 2b.

Next, we derive the y-parameters of the intrinsic part of the equivalent circuit in Fig. 2b, which is inside the blue frame. We have considered such equivalent circuit as a two-port network connected in a common source configuration. The intrinsic Y-parameters ($Y_i$) can be written as:

$$Y_i(\omega) = \begin{pmatrix} y_{11,i} & y_{12,i} \\ y_{21,i} & y_{22,i} \end{pmatrix} \begin{cases} y_{11,i} = i(C_{gd} + C_{gs})\omega \\ y_{12,i} = -iC_{gd}\omega \\ y_{21,i} = g_m - iC_{dg}\omega \\ y_{22,i} = g_{ds} + i(C_{gd} + C_{sd})\omega \end{cases} \quad (1)$$

where $\omega = 2\pi f$ and $f$ is the frequency of the AC signal and ports 1 and 2 refer to the gate-source and drain-source ports, respectively.

Consequently, the Z-parameters of the equivalent circuit can be expressed as:

$$Z(\omega) = \left[ Y_i(\omega)^{-1} + R \right] \quad \text{where} \quad R = \begin{pmatrix} R_g + R_s & R_s \\ R_s & R_d + R_s \end{pmatrix} \quad (2)$$

### B. RF performance of 2D-FETs

Whenever investigating a new technology for electronic applications, it is of primary importance to get the figures of merit (FoMs) and compare them against the requirements of the International Technology Roadmap for Semiconductors (ITRS). Considering the target of high frequency electronics, the cut-off frequency ($f_{Tx}$) and the maximum oscillation frequency ($f_{max}$) are the most widely used FoMs. The cut-off frequency is defined as the frequency for which the magnitude of the small-signal current gain ($h_{21}$) of the transistor is reduced to unity:

$$h_{21}(\omega) = -\frac{y_{21}}{y_{11}} \rightarrow |h_{21}(2\pi f_{Tx})| = 1 \quad (3)$$

where the y-parameters entering in (3) come from the impedance matrix calculated in (2):

$$Y(\omega) = \begin{pmatrix} y_{11} & y_{12} \\ y_{21} & y_{22} \end{pmatrix} = Z(\omega)^{-1} \quad (4)$$

On the other hand, the maximum oscillation frequency ($f_{max}$) is defined as the highest possible frequency for which the magnitude of the power gain ($U$, Mason's invariant) of the transistor is reduced to unity.

$$U(\omega) = \frac{|y_{12} - y_{21}|^2}{4(\text{Re}[y_{11}]\text{Re}[y_{22}] - \text{Re}[y_{12}]\text{Re}[y_{21}])}; \ U(2\pi f_{max}) = 1 \quad (5)$$

We have found significant discrepancies between our model and other models regarding the evaluation of the RF FoMs. The reasons for that are the following: (i) the reported expressions have been obtained after assuming a small-signal equivalent circuit based on the Meyer-like capacitance approach (which always assume capacitance reciprocity), similar as the one depicted in Fig. 2a; and (ii) approximations usually made for conventional technologies as, for example, if the transistor is working in the saturation region, then, the drain edge of the device is depleted of mobile charge carriers, so $C_{gd}$ can be neglected with respect to $C_{gs}$. So, in order to keep the accuracy in evaluating the FoMs to the highest level, we have obtained new explicit expressions with no approximations to compute the RF FoMs based on the equivalent circuit presented in Fig. 2b. In doing so, the definitions of both $f_{Tx}$ and $f_{max}$ given by (3) and (5) have been applied to obtain (7) and (9), respectively. Explicit expressions for the intrinsic RF FoMs have also been provided in (6) and (8), respectively, considering $R_s = R_d = 0$.

$$f_{T,i} = \frac{|g_m|}{2\pi\sqrt{(C_{gs}+C_{gd})^2 - C_{dg}^2}} \tag{6}$$

$$\begin{aligned} f_{Tx} &= \frac{\sqrt{-(c_1+c_2)}}{2\pi\sqrt{c_3}} \\ C_{gg} &= C_{gd} + C_{gs} \\ c_1 &= -C_{dg}^2 + \left(C_{gg} + (C_{gg}g_{ds}+C_{gd}g_m)R_d\right)^2 + 2\left(C_{gg}\left((-C_{dg}+C_{gg})g_{ds} - C_{sd}g_m\right) + (C_{gg}g_{ds}+C_{gd}g_m)^2 R_d\right)R_s \\ c_2 &= -\sqrt{2c_3 g_m^2 + \left(-C_{dg}^2 + (C_{gg}+(C_{gg}g_{ds}+C_{gd}g_m)R_d)^2 + 2(C_{gg}((-C_{dg}+C_{gg})g_{ds}-C_{sd}g_m)+(C_{gg}g_{ds}+C_{gd}g_m)^2 R_d)R_s\right)^2} \\ c_3 &= 2\left(C_{dg}C_{gd} - C_{gg}(C_{gd}+C_{sd})\right)^2 R_d(R_d+2R_s) \end{aligned} \tag{7}$$

$$f_{max,i} = \frac{|g_m|}{2\pi\sqrt{4R_g(C_{gd}+C_{gs})(C_{gs}g_{ds}+C_{gd}(g_{ds}+g_m)) - (C_{dg}-C_{gd})^2}} \tag{8}$$

$$\begin{aligned} f_{max} &= \frac{\sqrt{-(a_1+a_2)}}{4\pi\sqrt{a_3}} \\ C_{gg} &= C_{gd}+C_{gs} \qquad\qquad R_{gds} = R_g R_d + R_d R_s + R_g R_s \\ a_1 &= -C_{dg}^2 + 2C_{dg}\left(C_{gd}+2g_{ds}(C_{gd}R_d - C_{gs}R_s)\right) + b_1 + b_2 + b_3 \qquad a_2 = -\sqrt{8a_3 g_m^2 + a_1^2} \\ a_3 &= 2\left(C_{dg}C_{gd} - C_{gg}(C_{gd}+C_{sd})\right)^2 R_{gds} \qquad b_1 = 4C_{gd}\left(C_{sd}g_m R_d + C_{gs}\left(g_m R_g + 2g_{ds}^2 R_{gds} + g_{ds}(2R_g + 2g_m R_{gds} + R_s)\right)\right) \\ b_2 &= 4C_{gs}\left(-C_{sd}g_m R_s + C_{gs}g_{ds}(R_g+R_s+g_{ds}R_{gds})\right) \qquad b_3 = C_{gd}^2\left(-1 + 4g_{ds}R_g + 4\left(g_m(R_d+R_g) + (g_{ds}+g_m)^2 R_{gds}\right)\right) \end{aligned} \tag{9}$$

$$R_c = \frac{\text{Re}(z_{22})\text{Im}(z_{12}) - \text{Re}(z_{12})\text{Im}(z_{22})}{2\text{Im}(z_{12}) - \text{Im}(z_{22})} \tag{10}$$

$$g_m = \frac{(2\text{Im}[z_{12}] - \text{Im}[z_{22}])(2\text{Im}[z_{21}]\text{Re}[z_{12}] - \text{Im}[z_{22}]\text{Re}[z_{12}] - 2\text{Im}[z_{12}]\text{Re}[z_{21}] + \text{Im}[z_{22}]\text{Re}[z_{21}] + \text{Im}[z_{12}]\text{Re}[z_{22}] - \text{Im}[z_{21}]\text{Re}[z_{22}])}{(\text{Im}[z_{12}]\text{Im}[z_{21}] - \text{Im}[z_{11}]\text{Im}[z_{22}])\left((-2\text{Im}[z_{12}] + \text{Im}[z_{22}])^2 + (-2\text{Re}[z_{12}] + \text{Re}[z_{22}])^2\right)} \tag{11}$$

$$g_{ds} = \frac{(2\text{Im}[z_{12}] - \text{Im}[z_{22}])\left(\text{Im}[z_{11}]\text{Im}[z_{22}](-2\text{Re}[z_{12}] + \text{Re}[z_{22}]) + \text{Im}[z_{12}]\left(\text{Im}[z_{22}]\text{Re}[z_{12}] + \text{Re}[z_{21}](2\text{Im}[z_{12}] - \text{Im}[z_{22}]) - \text{Im}[z_{12}]\text{Re}[z_{22}]\right)\right)}{\text{Im}[z_{22}](\text{Im}[z_{12}]\text{Im}[z_{21}] - \text{Im}[z_{11}]\text{Im}[z_{22}])\left((-2\text{Im}[z_{12}] + \text{Im}[z_{22}])^2 + (-2\text{Re}[z_{12}] + \text{Re}[z_{22}])^2\right)} \tag{12}$$

$$R_g = \frac{\text{Im}[z_{22}]^2(-\text{Re}[z_{11}] + \text{Re}[z_{12}]) + \text{Im}[z_{12}]\text{Im}[z_{22}](2\text{Re}[z_{11}] - \text{Re}[z_{21}] - \text{Re}[z_{22}]) + \text{Im}[z_{12}]^2(-2\text{Re}[z_{21}] + \text{Re}[z_{22}])}{(2\text{Im}[z_{12}] - \text{Im}[z_{22}])\text{Im}[z_{22}]} \tag{13}$$

$$C_{gs} = -\frac{\text{Im}[z_{12}] - \text{Im}[z_{22}]}{\omega(\text{Im}[z_{12}]\text{Im}[z_{21}] - \text{Im}[z_{11}]\text{Im}[z_{22}])} \tag{14}$$

$$C_{gd} = C_{gs}\frac{\text{Im}(z_{12})}{\text{Im}(z_{22}) - \text{Im}(z_{12})} = \frac{\text{Im}[z_{12}]}{\omega(\text{Im}[z_{12}]\text{Im}[z_{21}] - \text{Im}[z_{11}]\text{Im}[z_{22}])} \tag{15}$$

$$\begin{aligned} C_{dg} &= \frac{4\text{Im}[z_{12}]^2 \text{Im}[z_{21}] + \text{Im}[z_{22}]\left(\text{Im}[z_{21}]\text{Im}[z_{22}] + (\text{Re}[z_{12}] - \text{Re}[z_{21}])(2\text{Re}[z_{12}] - \text{Re}[z_{22}])\right)}{(\text{Im}[z_{12}]\text{Im}[z_{21}] - \text{Im}[z_{11}]\text{Im}[z_{22}])\left((-2\text{Im}[z_{12}] + \text{Im}[z_{22}])^2 + (-2\text{Re}[z_{12}] + \text{Re}[z_{22}])^2\right)\omega} \\ &+ \frac{\text{Im}[z_{12}]\left(4(\text{Re}[z_{12}]\text{Re}[z_{21}] - \text{Im}[z_{21}]\text{Im}[z_{22}]) - 2(\text{Re}[z_{12}] + \text{Re}[z_{21}])\text{Re}[z_{22}] + \text{Re}[z_{22}]^2\right)}{(\text{Im}[z_{12}]\text{Im}[z_{21}] - \text{Im}[z_{11}]\text{Im}[z_{22}])\left((-2\text{Im}[z_{12}] + \text{Im}[z_{22}])^2 + (-2\text{Re}[z_{12}] + \text{Re}[z_{22}])^2\right)\omega} \end{aligned} \tag{16}$$

$$\begin{aligned} C_{sd} &= \frac{-\text{Im}[z_{12}]\left(\text{Im}[z_{22}]\left(\text{Im}[z_{22}]^2 + (2\text{Re}[z_{12}] - \text{Re}[z_{22}])(\text{Re}[z_{12}] + \text{Re}[z_{21}] - \text{Re}[z_{22}])\right) + \text{Im}[z_{12}]\left(-4\text{Im}[z_{22}]^2 + (2\text{Re}[z_{12}] - \text{Re}[z_{22}])(-2\text{Re}[z_{21}] + \text{Re}[z_{22}])\right)\right)}{\text{Im}[z_{22}](\text{Im}[z_{12}]\text{Im}[z_{21}] - \text{Im}[z_{11}]\text{Im}[z_{22}])\left((-2\text{Im}[z_{12}] + \text{Im}[z_{22}])^2 + (-2\text{Re}[z_{12}] + \text{Re}[z_{22}])^2\right)\omega} \\ &+ \frac{\text{Im}[z_{11}]\text{Im}[z_{22}](-2\text{Im}[z_{12}] + \text{Im}[z_{22}])^2 - 4\text{Im}[z_{12}]^3 \text{Im}[z_{22}]}{\text{Im}[z_{22}](\text{Im}[z_{12}]\text{Im}[z_{21}] - \text{Im}[z_{11}]\text{Im}[z_{22}])\left((-2\text{Im}[z_{12}] + \text{Im}[z_{22}])^2 + (-2\text{Re}[z_{12}] + \text{Re}[z_{22}])^2\right)\omega} \end{aligned} \tag{17}$$

*C. Parameter extraction methodology*

A method for extracting small-signal parameters from S-parameter measurements has been reported for a charge-based model in the context of silicon technology [24]. However, it assumes that the metal contact and access resistances can be neglected, which is not the case in 2D-FETs, so this methodology cannot be directly applied without introducing large errors. The issue is that the common de-embedding procedures does not allow extracting those resistances [10], [25]–[29]. Instead, they should be extracted apart; for instance, by using the TLM.

The most common de-embedding procedure consists of applying "open" and "short" structures to identical layouts, one excluding the 2D-channel, so to remove the effect of the probing pads, metal interconnections, including the parasitic capacitances and parasitic inductances. Since the effect of the 2D-channel cannot be removed by the de-embedding process, the parasitic resistance extracted by this method do not include either the metal contact resistance or the access resistance [12]. Consequently, they should be included as a part of the small-signal equivalent circuit. It is worth noting that such a methodology proposed here is suitable for any FET with high contact and/or access resistances that could not be extracted separately.

In doing so, we have included the effect of them in the parameter extraction methodology, so they can be extracted together with the rest of intrinsic parameters from S-parameter measurements. The contact resistance with a 2D-material is currently an important bottleneck, together with the lack of perfect current saturation, hampering the realization of power gain at terahertz frequencies [30]–[32]. On the other hand, in many embodiments of the 2D-transistor an ungated area exists between the drain/source metal and the channel under the gate resulting in additional access resistance, which should be considered.

So, a suitable parameter extraction method should be as the one described in the following steps:

1) Apply "open" and "short" structures to identical device under test's layouts, one excluding the 2D-channel, in order to remove the effect of the probing pads including the parasitic capacitances and parasitic inductances [10],[25]–[29].

2) Extract the series combination of the metal contact and access resistances using equation (10), where we have assumed that both drain and source resistances are the same, namely: $R_s = R_d = R_c$. Other possibility to estimate these extrinsic resistances is relying on the TLM, which is the most common procedure.

3) Direct application of the equations (11)-(17) to obtain the transconductance ($g_m$), output conductance ($g_{ds}$), gate resistance ($R_g$) and the intrinsic capacitances ($C_{gs}$, $C_{gd}$, $C_{dg}$, $C_{sd}$). These expressions have been derived with no approximations.

As a matter of convenience we have expressed equations (10)-(17) in terms of the Z-parameters instead of S-parameters that we had announced. The equivalence between both kind of parameters is well known and can be found in [33]. It is important to highlight that the extraction approach above-mentioned allows to get the small-signal parameters at any arbitrary bias. This is in contrast to the extraction method reported in [12] that requires biasing the GFET at the minimum conductivity to extract the intrinsic capacitances. So, this procedure is fine when the model is operated close to the Dirac voltage, but discrepancies could arise far from this bias point according to the bias dependence of such intrinsic capacitances observed in Fig. 3c.

## III. RESULTS AND DISCUSSION

*A. Assessment of the RF performance calculation of GFETs*

In order to assess the new expressions (7) and (9) to estimate the RF FoMs, we have obtained the small-signal parameters of a prototype GFET described in Table I from the large-signal model presented in [17], [34]. The gate bias dependence of the transconductance and output conductance is depicted in Figs. 3a-b and 4a-b, for a drain bias $V_{DS} = 0.5$ V and $V_{DS} = 3$ V, respectively, the latter representative of the GFET biased in the NDR region. The intrinsic capacitances for $V_{DS} = 0.5$ V are shown in Fig. 3c. We have calculated $f_{Tx}$ and $f_{max}$ using different expressions found in the literature, specifically the ones provided in [10], [13], [35]–[37]. Results are presented in Figs. 3d-e and 4c-d.

TABLE I. INPUT PARAMETERS OF A PROTOTYPE GFET (LARGE-SIGNAL MODEL PRESENTED IN [17])

| Input parameter | Value | Input parameter | Value |
|---|---|---|---|
| $T$ | 300 K | $L$ | 1 μm |
| $\mu$ | 2000 cm$^2$/Vs | $W$ | 10 μm |
| $V_{gs0}$ | 0 V | $L_t$ | 12 nm |
| $\Delta$ | 0.08 eV | $\varepsilon_{top}$ | 9 |
| $R_s$, $R_d$ | 200 Ω·μm | $R_g$ | 5 Ω·μm |

Both $f_{Tx}$ and $f_{max}$ expressions from [10], [35], [37] can largely underestimate or overestimate the values depending on the gate voltage overdrive. However, results from [36] are far and, in particular, for $V_{DS} = 3$ V there is a gate bias region where the $f_{Tx}$ and $f_{max}$ expression results in imaginary or real negative values. Regarding $f_{max}$ evaluation we have assessed the case where a GFET is operated in its NDR region, which is a feature of interest in many applications [37]–[42]. As suggested in Fig. 4d, there is no expression found in the literature which gives a positive real estimation within this gate bias range. The model we are proposing is an exception, delivering results that are not imaginary or real negative.

Moreover, we have calculated the RF FoMs assuming a Meyer-like model as the one depicted in Fig. 2a, by enforcing $C_{dg} = C_{gd}$ and $C_{sd} = 0$ in equations (7) and (9). This has been done for the sake of highlighting the differences with the charge-based model. Results have been plotted in Figs. 3d-e and 4c-d (yellow lines). Especially in Fig.4c we can realize on the importance of assuming a charge-based model and consistently estimating the RF FoMs in accordance to it. In addition, for the sake of sensitivity evaluation, the partial derivatives of $f_{Tx}$ with respect to the extrinsic elements have been calculated for the DUT at $V_{DS} = 0.5$ V. Specifically, $\partial f_{Tx}/\partial R_d$ can be up to ~0.13 GHz/Ω while $\partial f_{Tx}/\partial R_s$ ~0.07 GHz/Ω.

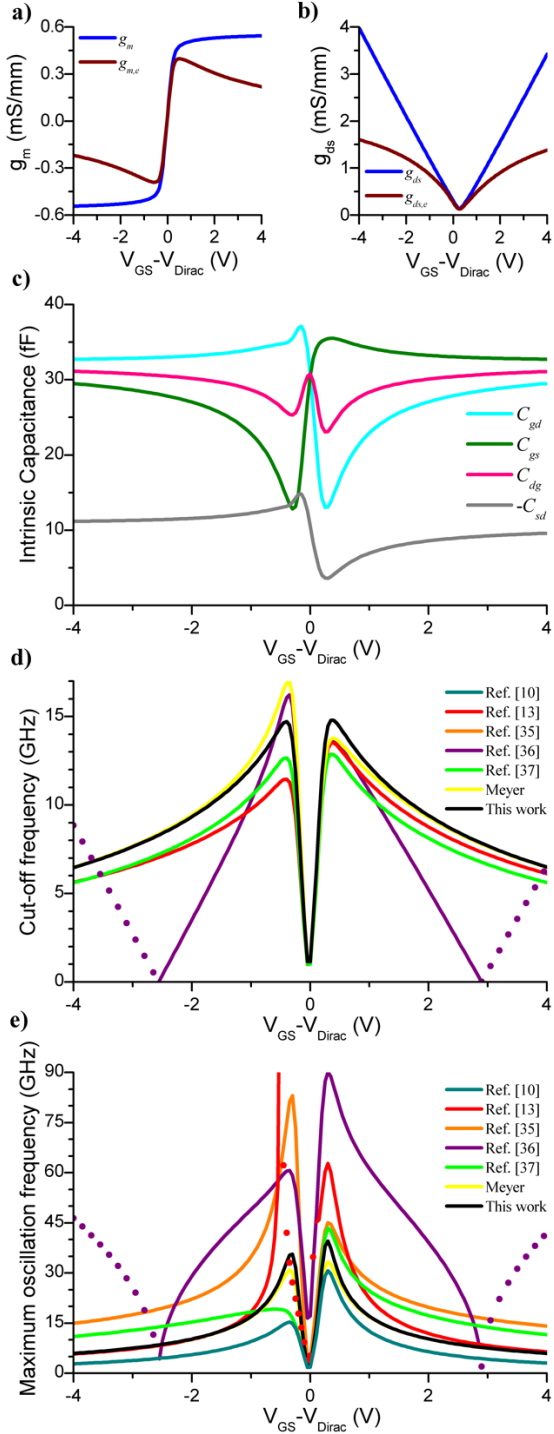

Fig. 3 Gate bias dependence of the small-signal parameters and RF FoMs of the GFET described in Table I for a drain bias $V_{DS} = 0.5$ V. The closed circles represent the absolute value of the frequency, where the calculated values are real negative or imaginary. a) Intrinsic ($g_m$) and extrinsic ($g_{m,e}$) transconductance; b) intrinsic ($g_{ds}$) and extrinsic ($g_{ds,e}$) output conductance; c) intrinsic capacitances ($C_{gd}$, $C_{gs}$, $C_{dg}$, $C_{sd}$); d) cut-off frequency ($f_{Tx}$); and e) maximum oscillation frequency ($f_{max}$).

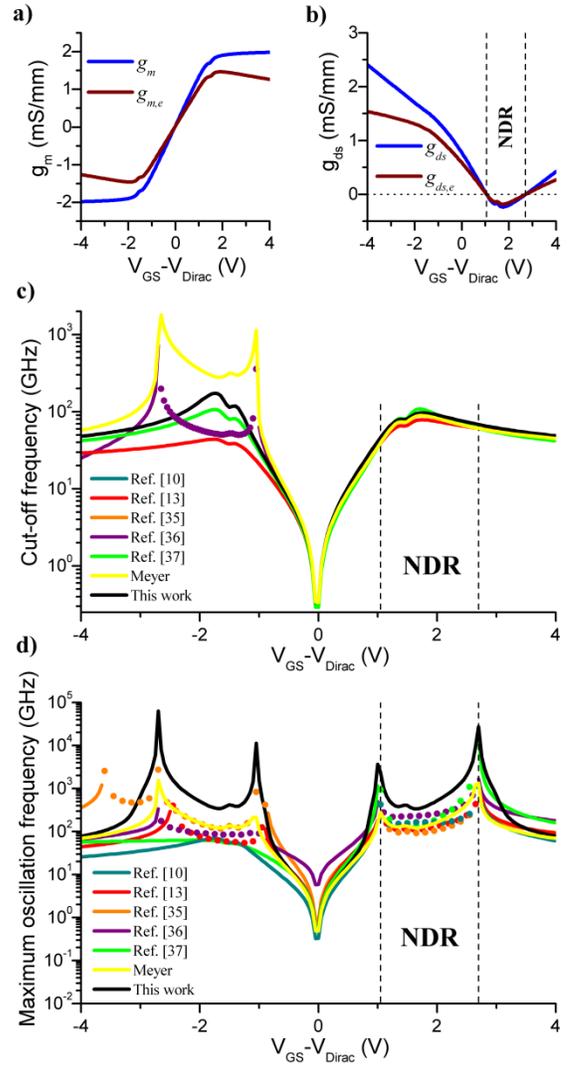

Fig. 4 Gate bias dependence of the small-signal parameters and RF FoMs of the GFET described in Table I for a drain bias $V_{DS} = 3$ V. The closed circles represent the absolute value of the frequency, where the calculated values are real negative or imaginary. a) Intrinsic ($g_m$) and extrinsic ($g_{m,e}$) transconductance; b) intrinsic ($g_{ds}$) and extrinsic ($g_{ds,e}$) output conductance. Notice that there is a region of negative differential resistance (NDR) in the range of $V_{GS} = [1.05 – 2.7]$ V; c) cut-off frequency ($f_{Tx}$); and d) maximum oscillation frequency ($f_{max}$).

### B. Extracting the small-signal parameters of a GFET

To assess the proposed parameter extraction method, a state-of-the-art GFET has been characterized in both DC and RF. A SEM image of the GFET ($W = 12$ μm, $L = 100$ nm) is shown in Fig. 1b and its fabrication process has been described in [43].

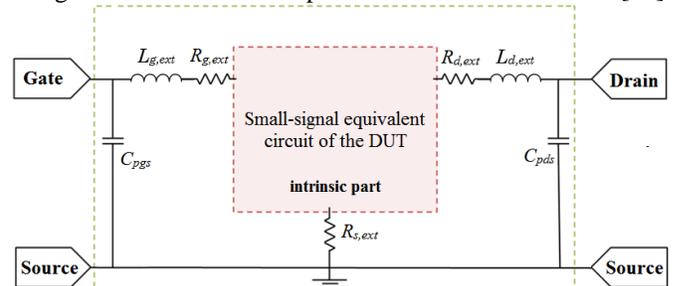

Fig. 5 Topology of the small-signal equivalent circuit of the microwave GFET under test including extrinsic elements. The intrinsic part could be either of the networks depicted in Fig. 2 depending on the capacitance model considered.

TABLE II. EXTRACTED EXTRINSIC ELEMENTS OF THE GFET DESCRIBED IN [43] AFTER DE-EMBEDDING
($V_{GS,e}$ = 0.2 V and $V_{DS,e}$ = 1 V)

| Element | Value | Element | Value |
|---|---|---|---|
| $R_{g,ext}$ | 42 Ω | $C_{pgs}$ | 12 fF |
| $R_{d,ext}$ | 110 Ω | $C_{pds}$ | 12 fF |
| $R_{s,ext}$ | 110 Ω | $L_{g,ext} = L_{d,ext}$ | 0 nH |

TABLE III. EXTRACTED SMALL-SIGNAL PARAMETERS OF THE CHARGE-BASED MODEL FOR THE EXEMPLARY GFET FROM [43]
($V_{GS,e}$ = 0.2 V and $V_{DS,e}$ = 1 V)

| Element | Value | Element | Value |
|---|---|---|---|
| $C_{gs}$ | 6.5 fF | $g_m$ | 1.55 mS |
| $C_{gd}$ | 9.5 fF | $g_{ds}$ | -6.5 mS |
| $C_{dg}$ | 10.5 fF | $R_g$ | 0.5 Ω |
| $C_{sd}$ | -3.5 fF | $R_d = R_s$ | 215 Ω |

The high-frequency performance of the GFET was characterized using a Vector Network Analyzer (Agilent, E8361A) under ambient conditions in the frequency range of 0.25 – 45 GHz. A common calibration procedure of line-reflect-reflect-match was performed before measurements. Fig. 5 shows the topology of the small-signal equivalent circuit for the microwave GFET under test including the extrinsic elements. Those elements represent the contributions arising from the interconnections between the device and the outside. The de-embedding procedure was implemented to subtract the unwanted contribution of such an extrinsic network, as described in [28], [29], [44]. The values of the extrinsic elements of the DUT are given in Table II. However, the effect of the series combination of the drain/source contact and access resistances could not be de-embedded by the open and short test structures. Following the extraction method described in section II.C, the intrinsic small-signal parameters have been obtained and summarized in Table III. We have checked that the extracted parameters are insensitive to the frequency selected for getting the S-parameters. Also notice that, due to the non-reciprocity, $C_{dg}$ and $C_{gd}$ are different. Besides, measured and modeled S-parameters at $V_{GS,e}$ = 0.2 V and $V_{DS,e}$ = 1 V plotted together in Fig. 6 are in good agreement. For the sake of completeness, the bias dependence of the extracted model parameters, as well as the corresponding RF FoMs of the GFET, can be found in the appendix.

The extracted value of the series resistance $R_c = R_s = R_d$ = 215 Ω is in good agreement with the average contact resistance reported (around 2200 Ω·μm) for the devices fabricated in [43]. Notice the importance of considering the extraction of these non-negligible resistances after the de-embedding procedure when modeling 2D-FETs.

On the other hand, we can calculate the extrinsic transconductance ($g_{m,e}$) and the extrinsic output conductance ($g_{ds,e}$) as following [45]:

$$g_{m,e} = \frac{\partial I_{DS}}{\partial V_{GS,e}} = \frac{g_m}{1 + g_m R_s + g_{ds}(R_s + R_d)}$$
$$g_{ds,e} = \frac{\partial I_{DS}}{\partial V_{DS,e}} = \frac{g_{ds}}{1 + g_m R_s + g_{ds}(R_s + R_d)} \quad (18)$$

In [43], a $g_{m,e}$ of ~ -100 μS/μm and a $g_{ds,e}$ of ~ 370 μS/μm were reported at $V_{GS,e}$ = 0.2 V and $V_{DS,e}$ = 1 V. They were extracted from the DC transfer characteristics ($I_{DS}$ vs. $V_{GS,e}$ curve) and from the output characteristics ($I_{DS}$ vs. $V_{DS,e}$ curve), respectively. These values are in good agreement with the ones calculated by equation (18), using the parameters in Table III, which have been obtained following the parameter extraction methodology explained before.

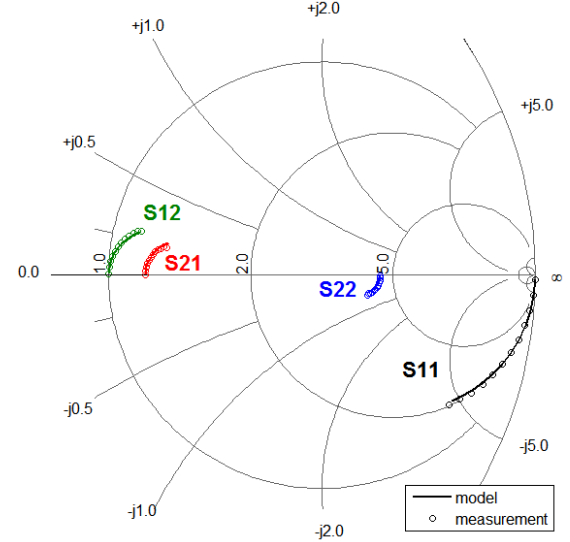

Fig. 6 S-parameter measurements (circles) and simulations (lines) for the applied bias $V_{GS,e}$ = 0.2 V and $V_{DS,e}$ = 1 V.

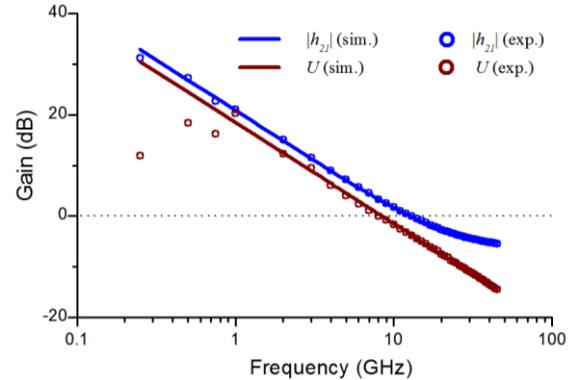

Fig. 7 Radio-frequency performance of the GFET characterized in Fig. 6 ($V_{GS,e}$ = 0.2 V and $V_{DS,e}$ = 1 V) with parameters listed in Table III. Measured (symbols) and simulated (solid line) small-signal current gain ($|h_{21}|$) and Mason's invariant ($U$) plotted versus frequency.

TABLE IV. ESTIMATION OF THE RF FOMS OF THE GFET FROM [43]
(imaginary values are written in italic style)

| | $f_{Tx}$ [GHz] | $f_{max}$ [GHz] |
|---|---|---|
| This work | 11.92 | 8.59 |
| Ref. [10] | -11.02 | *4.65* |
| Ref. [13] | 13.69 | 6.75 |
| Ref. [35] | -11.02 | *-16.04* |
| Ref. [36] | -11.89 | 315.65 |
| Ref. [37] | -11.02 | -25.45 |

Finally, Fig. 7 shows the experimental current gain ($|h_{21}|$) and Mason's invariant ($U$), both obtained from the S-parameter measurements depicted in Fig. 6, compared to the simulated ones obtained from the small-signal model. Both $f_{Tx}$ and $f_{max}$ coming from different approaches have been calculated using the extracted parameters listed in Table III. They have been summarized in Table IV, showing a large dispersion of values, being the values from (7) and (9) the more accurate prediction. Notice that, because of the negative intrinsic output conductance, many reported formulas give real negative or imaginary values.

## IV. Conclusions

A small-signal model for three-terminal 2D-FETs has been presented. The model formulation is universally valid for 2D-materials such as graphene and 2D-semiconductors. Two main features must be highlighted: (i) the small-signal model guarantees charge conservation and takes into account non-reciprocal capacitances and (ii) the metal contact and access resistances have been included in the parameter extraction methodology because of the impossibility of removing their effect from a de-embedding procedure.

Explicit and exact expressions for both cut-off and maximum oscillation frequency have been provided consistent with the charge-based small-signal model with no approximations. Such expressions have been compared with others found in the literature. We have found noticeable discrepancies among them when applied to GFETs, especially when the transistor is operated in the NDR region.

An approach to extract the small-signal parameters (transconductance, output conductance and intrinsic capacitances) and gate resistance from S-parameter measurements has been proposed. Additionally, direct extraction method of the series combination of the metal contact and access resistances from S-parameter measurements has also been provided. The extraction approach has been assessed against S-parameter measurements of a GFET in the RF regime, showing good agreement.

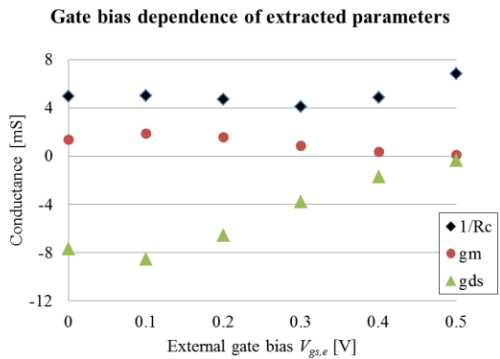

Fig. 8 Bias dependence of the extracted transconductance, output conductance and series combination of the contact and access resistance with the external gate bias for a fixed $V_{DS,e}$ = 1 V.

A charge-based small-signal model is important not only to ensure the model accuracy to predict the figures of merit but also to guarantee the compatibility with physics-based large-signal models. Moreover, charge conservation could also be critical when a large-signal model is assembled building up on small-signal models, in form of tables containing values of drain current and of small-signal parameters for many combinations of bias voltages. Such a model is the so-called table look-up model. Then, by using interpolation functions the values for points in between could be computed.

## Appendix

In order to examine the bias dependence of the RF FoMs, we have extracted the small-signal parameters of the DUT introduced in section III.B for an external gate bias $V_{GS,e}$ ranging from 0 to 0.5 V while keeping a constant external drain bias of $V_{DS,e}$ = 1V. The results have been shown in Figs. 8-9. With this information, the bias dependence of the RF FoMs can be calculated using equations (7) and (9), and the result has been plotted in Fig. 10.

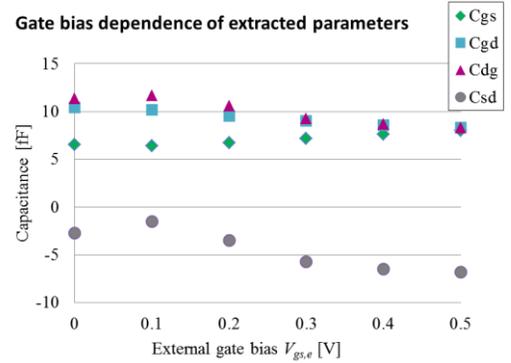

Fig. 9 Bias dependence of the extracted intrinsic capacitances with the external gate bias for a fixed $V_{DS,e}$ = 1 V.

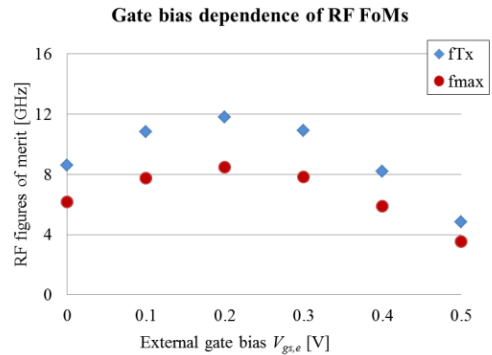

Fig. 10 Bias dependence of the RF figures of merit with the external gate bias for a fixed $V_{DS,e}$ = 1 V.

## References

[1] A. C. Ferrari, F. Bonaccorso, V. Falko, K. S. Novoselov, S. Roche, P. Bøggild, S. Borini, F. Koppens, V. Palermo, N. Pugno, J. a. Garrido, R. Sordan, A. Bianco, L. Ballerini, M. Prato, E. Lidorikis, J. Kivioja, C. Marinelli, T. Ryhänen, A. Morpurgo, J. N. Coleman, V. Nicolosi, L. Colombo, A. Fert, M. Garcia-Hernandez, A. Bachtold, G. F. Schneider, F. Guinea, C. Dekker, M. Barbone, C. Galiotis, A. Grigorenko, G. Konstantatos, A. Kis, M. Katsnelson, C. W. J. Beenakker, L. Vandersypen, A. Loiseau, V. Morandi, D. Neumaier, E. Treossi, V. Pellegrini, M. Polini, A. Tredicucci, G. M. Williams, B. H. Hong, J. H. Ahn, J. M. Kim, H. Zirath, B. J. van Wees, H. van der Zant, L. Occhipinti, A. Di Matteo, I. a. Kinloch, T. Seyller, E. Quesnel, X. Feng, K. Teo, N. Rupesinghe, P. Hakonen, S. R. T. Neil, Q. Tannock, T. Löfwander, and J. Kinaret, "Science and technology roadmap for graphene, related two-dimensional crystals, and hybrid systems," *Nanoscale*, vol. 7, no. 11, pp. 4598–4810, Mar. 2015.

[2] F. Schwierz, R. Granzner, and J. Pezoldt, "Two-dimensional materials and their prospects in transistor electronics," *Nanoscale*, pp. 8261–8283, 2015.

[3] G. Fiori, F. Bonaccorso, G. Iannaccone, T. Palacios, D. Neumaier, A. Seabaugh, S. K. Banerjee, and L. Colombo, "Electronics based on two-dimensional materials.," *Nat. Nanotechnol.*, vol. 9, no. 10, pp. 768–779, Oct. 2014.

[4] M. Chhowalla, "Two-dimensional semiconductors for transistors," *Nat. Rev. Mater.*, vol. 1, no. 11, p. 16052, Aug. 2016.

[5] W. Zhu, S. Park, M. N. Yogeesh, K. M. McNicholas, S. R. Bank, and D. Akinwande, "Black Phosphorus Flexible Thin Film Transistors at Gighertz Frequencies," *Nano Lett.*, vol. 16, no. 4, pp. 2301–2306, Apr. 2016.

[6] S. Park, W. Zhu, H.-Y. Chang, M. N. Yogeesh, R. Ghosh, S. K. Banerjee, and D. Akinwande, "High-frequency prospects of 2D nanomaterials for flexible nanoelectronics from baseband to sub-THz devices," in *2015 IEEE International Electron Devices Meeting (IEDM)*, 2015, p. 32.1.1-32.1.4.

[7] C. Yan, J. H. Cho, and J.-H. Ahn, "Graphene-based flexible and stretchable thin film transistors," *Nanoscale*, vol. 4, no. 16, p. 4870, 2012.

[8] P. J. Jeon, J. S. Kim, J. Y. Lim, Y. Cho, A. Pezeshki, H. S. Lee, S. Yu, S.-W. Min, and S. Im, "Low Power Consumption Complementary Inverters with n-MoS2 and p-WSe2 Dichalcogenide Nanosheets on Glass for Logic and Light-Emitting Diode Circuits," *ACS Appl. Mater. Interfaces*, vol. 7, no. 40, pp. 22333–22340, Oct. 2015.

[9] R. Cheng, J. Bai, L. Liao, H. Zhou, Y. Chen, L. Liu, Y.-C. Lin, S. Jiang, Y. Huang, and X. Duan, "High-frequency self-aligned graphene transistors with transferred gate stacks.," *Proc. Natl. Acad. Sci. U. S. A.*, vol. 109, no. 29, pp. 11588–92, Jul. 2012.

[10] D. Krasnozhon, D. Lembke, C. Nyffeler, Y. Leblebici, and A. Kis, "MoS2 transistors operating at gigahertz frequencies.," *Nano Lett.*, vol. 14, no. 10, pp. 5905–11, Oct. 2014.

[11] H. Wang, X. Wang, F. Xia, L. Wang, H. Jiang, Q. Xia, M. L. Chin, M. Dubey, and S. Han, "Black Phosphorus Radio-Frequency Transistors," *Nano Lett.*, vol. 14, no. 11, pp. 6424–6429, Nov. 2014.

[12] O. Habibpour, J. Vukusic, and J. Stake, "A Large-Signal Graphene FET Model," *IEEE Trans. Electron Devices*, vol. 59, no. 4, pp. 968–975, Apr. 2012.

[13] J. G. Champlain, "A physics-based, small-signal model for graphene field effect transistors," *Solid. State. Electron.*, vol. 67, no. 1, pp. 53–62, Jan. 2012.

[14] G. I. Zebrev, A. A. Tselykovskiy, D. K. Batmanova, and E. V. Melnik, "Small-Signal Capacitance and Current Parameter Modeling in Large-Scale High-Frequency Graphene Field-Effect Transistors," *IEEE Trans. Electron Devices*, vol. 60, no. 6, pp. 1799–1806, Jun. 2013.

[15] F. Xia, V. Perebeinos, Y. Lin, Y. Wu, and P. Avouris, "The origins and limits of metal-graphene junction resistance.," *Nat. Nanotechnol.*, vol. 6, no. 3, pp. 179–184, Mar. 2011.

[16] Y. Tsividis, *Operation and modeling of the MOS transistor*, 2nd ed. New York ; Oxford : Oxford University Press, 1999.

[17] F. Pasadas and D. Jiménez, "Large-Signal Model of Graphene Field-Effect Transistors—Part I: Compact Modeling of GFET Intrinsic Capacitances," *IEEE Trans. Electron Devices*, vol. 63, no. 7, pp. 2936–2941, Jul. 2016.

[18] D. E. Ward and R. W. Dutton, "A charge-oriented model for MOS transistor capacitances," *IEEE J. Solid-State Circuits*, vol. 13, no. 5, pp. 703–708, Oct. 1978.

[19] D. E. Ward, "Charge-based modeling of capacitance in MOS transistors," Stanford Univ., CA., 1981.

[20] Ping Yang, B. D. Epler, and P. K. Chatterjee, "An Investigation of the Charge Conservation Problem for MOSFET Circuit Simulation," *IEEE J. Solid-State Circuits*, vol. 18, no. 1, pp. 128–138, 1983.

[21] H.-J. Park, P. K. Ko, and C. Hu, "A charge conserving non-quasi-state (NQS) MOSFET model for SPICE transient analysis," *IEEE Trans. Comput. Des. Integr. Circuits Syst.*, vol. 10, no. 5, pp. 629–642, May 1991.

[22] O. C. G. Filho, A. I. A. Cunha, M. C. Schneider, and C. Galup-Montoro, "A compact charge-based MOSFET model for circuit simulation," in *1998 IEEE International Conference on Electronics, Circuits and Systems. Surfing the Waves of Science and Technology (Cat. No.98EX196)*, vol. 1, pp. 491–494.

[23] M. Jamal Deen and T. A. Fjeldly, *CMOS RF modeling, characterization and applications*. River Edge, N.J. : World Scientific, 2002.

[24] M. Je, I. Kwon, J. Han, H. Shin, and K. K. Lee, "CMOS RF modeling and parameter extraction approaches taking charge conservation into account," *2002 Int. Conf. Model. Simul. Microsystems - MSM 2002*, pp. 698–701, 2002.

[25] L. Liao, Y.-C. Lin, M. Bao, R. Cheng, J. Bai, Y. Liu, Y. Qu, K. L. Wang, Y. Huang, and X. Duan, "High-speed graphene transistors with a self-aligned nanowire gate," *Nature*, vol. 467, no. 7313, pp. 305–308, Sep. 2010.

[26] A. Sanne, R. Ghosh, A. Rai, M. N. Yogeesh, S. H. Shin, A. Sharma, K. Jarvis, L. Mathew, R. Rao, D. Akinwande, and S. Banerjee, "Radio Frequency Transistors and Circuits Based on CVD $MoS_2$," *Nano Lett.*, vol. 15, no. 8, pp. 5039–5045, Aug. 2015.

[27] Y. Che, Y.-C. Lin, P. Kim, and C. Zhou, "T-Gate Aligned Nanotube Radio Frequency Transistors and Circuits with Superior Performance," *ACS Nano*, vol. 7, no. 5, pp. 4343–4350, May 2013.

[28] L. Nougaret, H. Happy, G. Dambrine, V. Derycke, J.-P. Bourgoin, A. A. Green, and M. C. Hersam, "80 GHz field-effect transistors produced using high purity semiconducting single-walled carbon nanotubes," *Appl. Phys. Lett.*, vol. 94, no. 24, p. 243505, Jun. 2009.

[29] N. Meng, J. F. Fernandez, D. Vignaud, G. Dambrine, and H. Happy, "Fabrication and Characterization of an Epitaxial Graphene Nanoribbon-Based Field-Effect Transistor," *IEEE Trans. Electron Devices*, vol. 58, no. 6, pp. 1594–1596, Jun. 2011.

[30] S. Das, H.-Y. Chen, A. V. Penumatcha, and J. Appenzeller, "High Performance Multilayer MoS2 Transistors with Scandium Contacts," *Nano Lett.*, vol. 13, no. 1, pp. 100–105, Jan. 2013.

[31] F. A. Chaves, D. Jiménez, A. A. Sagade, W. Kim, J. Riikonen, H. Lipsanen, and D. Neumaier, "A physics-based model of gate-tunable metal–graphene contact resistance benchmarked against experimental data," *2D Mater.*, vol. 2, no. 2, p. 25006, 2015.

[32] Y. Du, H. Liu, Y. Deng, and P. D. Ye, "Device Perspective for Black Phosphorus Field-Effect Transistors: Contact Resistance, Ambipolar Behavior, and Scaling," *ACS Nano*, vol. 8, no. 10, pp. 10035–10042, Oct. 2014.

[33] D. M. Pozar, *Microwave Engineering Fourth Edition*. 2005.

[34] F. Pasadas and D. Jiménez, "Large-Signal Model of Graphene Field- Effect Transistors—Part II: Circuit Performance Benchmarking," *IEEE Trans. Electron Devices*, vol. 63, no. 7, pp. 2942–2947, Jul. 2016.

[35] F. Schwierz, "Graphene Transistors: Status, Prospects, and Problems," *Proc. IEEE*, vol. 101, no. 7, pp. 1567–1584, Jul. 2013.

[36] T. H. Taur, Yuan and Ning, *Fundamentals of Modern VLSI Devices*, Second Edi. New York: Cambridge Univ Press, 2005.

[37] R. Grassi, A. Gnudi, V. Di Lecce, E. Gnani, S. Reggiani, and G. Baccarani, "Exploiting Negative Differential Resistance in Monolayer Graphene FETs for High Voltage Gains," *IEEE Trans. Electron Devices*, vol. 61, no. 2, pp. 617–624, Feb. 2014.

[38] A. Nogaret, "Negative differential conductance materials for flexible electronics," *J. Appl. Polym. Sci.*, vol. 131, no. 24, p. 40169(1)-40169(15), 2014.

[39] R. Grassi, A. Gnudi, V. Di Lecce, E. Gnani, S. Reggiani, and G. Baccarani, "Boosting the voltage gain of graphene FETs through a differential amplifier scheme with positive feedback," *Solid. State. Electron.*, vol. 100, pp. 54–60, 2014.

[40] Y. Wu, D. B. Farmer, W. Zhu, S.-J. Han, C. D. Dimitrakopoulos, A. A. Bol, P. Avouris, and Y.-M. Lin, "Three-terminal graphene negative differential resistance devices.," *ACS Nano*, vol. 6, no. 3, pp. 2610–6, Mar. 2012.

[41] P. Sharma, L. S. Bernard, A. Bazigos, A. Magrez, and A. M. Ionescu, "Room-Temperature Negative Differential Resistance in Graphene Field Effect Transistors: Experiments and Theory," *ACS Nano*, no. 1, pp. 620–625, 2015.

[42] P. C. Feijoo, D. Jiménez, and X. Cartoixà, "Short channel effects in graphene-based field effect transistors targeting radio-frequency applications," *2D Mater.*, vol. 3, no. 2, p. 25036, Jun. 2016.

[43] W. Wei, X. Zhou, G. Deokar, H. Kim, M. Belhaj, E. Galopin, E. Pallecchi, D. Vignaud, and H. Happy, "Graphene FETs with aluminum bottom-gate electrodes and its natural oxide as dielectrics," *IEEE Trans. Electron Devices*, vol. 62, no. 9, pp. 2769–2773, 2015.

[44] G. Crupi, D. M. M. P. Schreurs, and A. Caddemi, "A Clear-Cut Introduction to the De-embedding Concept: Less is More," in *Microwave De-embedding: From Theory to Applications*, 2013.

[45] M. Shur, *Physics of Semiconductor Devices*. Upper Saddle River, NJ, USA: Prentice-Hall, Inc., 1990.